\documentstyle[graphicx]{mn2e}

\newif\ifAMStwofonts
\title[WD+RG systems as the progenitors of SNe Ia]{WD+RG systems as the progenitors of Type Ia supernovae}
\author[B. Wang and Z. Han]
{B. Wang$^{\rm 1,2}$ \thanks{E-mail: wangbo@ynao.ac.cn}, and Z. Han $^{\rm 1}$  \\
$^1$ National Astronomical Observatories/Yunnan Observatory, the
Chinese Academy of Sciences, Kunming 650011, China\\
$^2$ Graduate School of the Chinese Academy of Sciences, Beijing
100049, China}

\begin{document}
%\date{Accepted. Received}
\pagerange{\pageref{firstpage}--\pageref{lastpage}} \pubyear{2010}
\maketitle

\label{firstpage}

\begin{abstract}
Type Ia supernovae (SNe Ia) play an important role in the study of
cosmic evolution, especially in cosmology. There are several
progenitor models for SNe Ia proposed in the past years. In this
paper, by considering the effect of accretion disk instability on
the evolution of white dwarf (WD) binaries, we performed detailed
binary evolution calculations for the WD + red-giant (RG) channel of
SNe Ia, in which a carbon--oxygen WD accretes material from a RG
star to increase its mass to the Chandrasekhar mass limit. According
to these calculations, we mapped out the initial and final
parameters for SNe Ia in the orbital period--secondary mass ($\log
P^{\rm i}-M^{\rm i}_2$) plane for various WD masses for this
channel. We discussed the influence of the variation of the duty
cycle value on the regions for producing SNe Ia. Similar to previous
studies, this work also indicates that the long-period dwarf novae
offer a possible ways for producing SNe Ia. Meanwhile, we find that
the surviving companion stars from this channel have a low mass
after SN explosion, which may provide a means for the formation of
the population of single low-mass WDs ($<$0.45$\,M_{\odot}$).

\end{abstract}


\section{Introduction}           %% first-level sections will be auto-capitalized
\label{sect:intro}

Type Ia Supernovae (SNe Ia) are excellent cosmological distance
indicators due to their high luminosities and remarkable uniformity.
They have been applied successfully in determining cosmological
parameters (e.g. $\Omega$ and $\Lambda$; Riess et al. 1998;
Perlmutter et al. 1999). However, several key issues related to the
nature of their progenitors and the physics of the explosion
mechanisms are still not well understood (Hillebrandt \& Niemeyer
2000; Podsiadlowski et al. 2008; Wang et al. 2008a), and no SN Ia
progenitor system has been conclusively identified from before the
explosion. It is generally believed that SNe Ia are thermonuclear
explosions of carbon--oxygen white dwarfs (CO WDs) in binaries (for
the review see Nomoto et al. 1997). Over the past few decades, two
families of SN Ia progenitor models have been proposed, i.e. the
double-degenerate (DD) and single-degenerate (SD) models. Of these
two models, the SD model is widely accepted at present. It is
suggested that the DD model, which involves the merger of two CO WDs
(Iben \& Tutukov 1984; Webbink 1984; Han 1998), likely leads to an
accretion-induced collapse rather than to an SN Ia (Nomoto \& Iben
1985). For the SD model, the companion is probably a main-sequence
(MS) star or a slightly evolved subgiant star (WD + MS channel), a
red-giant star (WD + RG channel), or an He star (WD + He star
channel) (e.g. Hachisu et al. 1996; Li $\&$ van den Heuvel 1997;
Langer et al. 2000; Han $\&$ Podsiadlowski 2004, 2006; Chen $\&$ Li
2007, 2009; Meng et al. 2009; L\"{u} et al. 2009; Wang et al.
2009a,b; Wang, Li $\&$ Han 2010). Note that, some recent
observations have indirectly suggested that at least some SNe Ia can
be produced by a variety of different progenitor systems (e.g. Patat
et al. 2007; Voss \& Nelemans 2008; Wang et al. 2008b; Justham et
al. 2009).

An explosion following the merger of two WDs would leave no remnant,
while the companion star in the SD model would survive and
potentially be identifiable. Tycho's supernova is a Galactic SN Ia.
Ruiz-Lapuente et al. (2004) find in the remnant region that Tycho G,
a star similar to the sun but with a lower gravity, moves at more
than three times the mean velocity of the stars there. They argued
that Tycho G could be the surviving companion of the supernova. Note
that there has been no conclusive proof yet that any individual
object is the surviving companion star of a SN Ia. It will be a
promising method to test SN Ia progenitor models by identifying
their surviving companions (e.g. Wang \& Han 2009).

The WD + RG channel is a possible ways to produce SNe Ia, and is
supported by some observations. It is suggested that, RS Oph and T
CrB, both recurrent novae are probable SN Ia progenitors and belong
to the WD + RG channel (e.g. Belczy$\acute{\rm n}$ski \&
Mikolajewska 1998; Hachisu et al. 1999a, 2007; Sokoloski et al.
2006). Meanwhile, by detecting Na I absorption lines with low
expansion velocities, Patat et al. (2007) suggested that the
companion of the progenitor of SN 2006X may be an early RG star.
Additionally, Voss \& Nelemans (2008) studied the pre-explosion
archival X-ray images at the position of the recent SN 2007on, and
they consider its progenitor as a WD + RG system.

Xu \& Li (2009) recently emphasized that the mass-transfer through
the Roche lobe overflow (RLOF) in the evolution of WD binaries may
become unstable (at least during part of the mass-transfer
lifetime). This important feature has been ignored in nearly all of
the previous theoretical works on SN Ia progenitors except for King
et al. (2003) \footnote{King et al. (2003) adopted a similar method
in Li \& Wang (1998) to produce SNe Ia with long period dwarf novae
in a semi-analytic approach.} and Xu \& Li (2009), who inferred that
the mass-accretion rate onto the WD during dwarf nova outbursts can
be sufficiently high to allow steady nuclear burning of the accreted
matter and growth of the WD mass. Following the work of Xu \& Li
(2009), Wang, Li \& Han (2010) studied the WD binaries towards SNe
Ia systematically. However, the interest of Wang, Li \& Han (2010)
mainly focused on the WD + MS channel of SNe Ia.

The purpose of this paper is to study the WD + RG channel towards
SNe Ia in a comprehensive manner, and to show the final parameter
spaces of companions at the moment of SN Ia explosion. In Section 2,
we simply describe the numerical code for the binary evolution
calculations. The binary evolutionary results are shown in Section
3. Finally, a discussion is given in Section 4.

\section{BINARY EVOLUTION CALCULATIONS}
In our WD binary models, the lobe-filling star is a RG star. The
star transfers some of its material onto the surface of the WD,
which increases the mass of the WD as a consequence. If the WD grows
to 1.378$\,M_{\odot}$, we assume that it explodes as an SN Ia. We
use Eggleton's stellar evolution code (Eggleton 1971, 1972, 1973) to
calculate the WD binary evolution. The code has been updated with
the latest input physics over the past three decades (Han et al.
1994; Pols et al. 1995, 1998). RLOF is treated within the code
described by Han et al. (2000). We set the ratio of mixing length to
local pressure scale height, $\alpha=l/H_{\rm p}$, to be 2.0 and set
the convective overshooting parameter, $\delta_{\rm OV}$, to be 0.12
(Pols et al. 1997; Schr$\ddot{\rm o}$der et al. 1997), which roughly
corresponds to an overshooting length of $\sim$0.25 pressure
scaleheights ($H_{\rm P}$). The opacity tables are compiled by Chen
\& Tout (2007). In our calculations, we use a typical Pop I
composition with H abundance $X=0.70$, He abundance $Y=0.28$ and
metallicity $Z=0.02$.

During the mass-transfer through the RLOF, the accreting material
can form an accretion disk surrounding the WD, which may become
thermally unstable when the effective temperature in the disk falls
below the H ionization temperature $\sim$6500\,K (e.g. van Paradijs
1996; King et al. 1997; Lasota 2001). This also corresponds to a
critical mass-transfer rate below which the disk is unstable.
Similar to the work of Xu \& Li (2009), we also set the critical
mass-transfer rate for a stable accretion disk to be
\begin{equation}
\dot{M}_{\rm cr, disk} \simeq 4.3\times 10^{-9}\,(P_{\rm orb}/4\,\rm
hr)^{1.7}\,\rm M_{\odot}\,yr^{-1},
\end{equation}
for WD accretors (van Paradijs 1996), where $P_{\rm orb}$ is the
orbital period. The locations of various types of cataclysmic
variable stars (e.g. the UX UMa, U Gem, SU UMa, and Z Cam systems)
in a ($P_{\rm orb}$, $\dot{M}_{\rm cr, disk}$) diagram are well
described by this expression (Smak 1983; Osaki 1996; van Paradijs
1996). If the mass-transfer rate, $|\dot{M}_2|$, is higher than the
critical value $\dot{M}_{\rm cr, disk}$, we assume that the
accretion disk is stable and the WD accretes smoothly at a rate
$\dot{M}_{\rm acc}=|\dot{M}_2|$; otherwise the WD accretes only
during outbursts and the mass-accretion rate is $\dot{M}_{\rm
acc}=|\dot{M}_2|/d$, where $d$ is the duty cycle. The mass-accretion
rate is $\dot{M}_{\rm acc}=0$ during quiescence. King et al. (2003)
showed that for typical values of the duty cycle $\sim$0.1 to a few
$10^{-3}$ the accretion rates onto the WD during dwarf nova
outbursts can be sufficiently high to allow steady nuclear burning
of the accreted matter. The limits on the duty cycles of dwarf nova
outbursts come from observations (Warner 1995): (1) The outburst
intervals for each object are quasi-periodic, but within the dwarf
nova family, the intervals can range from days to decades. (2) The
lifetime of an outburst is typically from 2 to 20 days and is
correlated with the outburst interval. The quasi-periodicity of the
dwarf nova outbursts allows to use a single duty cycle to roughly
describe the change in the mass-transfer rate, though we note that
this is a simplification of the real, complicated processes. Similar
to previous studies (e.g. Xu \& Li 2009), we also set the duty cycle
to be 0.01. Meanwhile, we also do some tests for a higher or lower
value of the duty cycle in our calculations.

Instead of solving stellar structure equations of a WD, we adopt the
prescription of Hachisu et al. (1999b) for the mass-growth of a CO
WD by accretion of H-rich material from its companion. The
prescription is given below. If the mass-accretion rate of the WD,
$\dot{M}_{\rm acc}$, is above a critical rate, $\dot{M}_{\rm
cr,WD}$, we assume that the accreted H steadily burns on the surface
of the WD and that the H-rich material is converted into He at a
rate $\dot{M}_{\rm cr,WD}$. The unprocessed matter is assumed to be
lost from the system as an optically thick wind at a mass-loss rate
$\dot{M}_{\rm wind}=|\dot{M}_{\rm 2}|-\dot{M}_{\rm cr,WD}$. The
critical mass-accretion rate is
 \begin{equation}
 \dot{M}_{\rm cr,WD}=5.3\times 10^{\rm -7}\frac{(1.7-X)}{X}(M_{\rm
 WD}/M_{\odot}-0.4) \,M_{\odot}\,yr^{-1},
  \end{equation}
where $X$ is the H mass fraction and $M_{\rm
 WD}$ is the mass of the accreting WD. The following assumptions are adopted when $|\dot{M}_{\rm acc}|$ is
smaller than $\dot{M}_{\rm cr,WD}$. (1) When $|\dot{M}_{\rm acc}|$
is less than $\dot{M}_{\rm cr,WD}$ but higher than
$\frac{1}{2}\dot{M}_{\rm cr,WD}$, the H-shell burning is steady and
no mass is lost from the system. (2) When $|\dot{M}_{\rm acc}|$ is
lower than $\frac{1}{2}\dot{M}_{\rm cr,WD}$ but higher than
$\frac{1}{8}\dot{M}_{\rm cr,WD}$, a very weak H-shell flash is
triggered but no mass is lost from the system. (3) When
$|\dot{M}_{\rm acc}|$ is lower than $\frac{1}{8}\dot{M}_{\rm
cr,WD}$, the H-shell flash is so strong that no material is
accumulated onto the surface of the WD.

We define the mass-growth rate of the He layer under the H-shell
burning as
 \begin{equation}
 \dot{M}_{\rm He}=\eta _{\rm H}|\dot{M}_{\rm acc}|,
  \end{equation}
where $\eta _{\rm H}$ is the mass-accumulation efficiency for
H-shell burning. According to the assumptions above, the values of
$\eta _{\rm H}$ are:
 \begin{equation}
\eta _{\rm H}=\left\{
 \begin{array}{ll}
 \dot{M}_{\rm cr,WD}/|\dot{M}_{\rm acc}|, & |\dot{M}_{\rm acc}|> \dot{M}_{\rm
 cr,WD},\\
 1, & \dot{M}_{\rm cr,WD}\geq |\dot{M}_{\rm acc}|\geq\frac{1}{8}\dot{M}_{\rm
 cr,WD},\\
 0, & |\dot{M}_{\rm acc}|< \frac{1}{8}\dot{M}_{\rm cr,WD}.
\end{array}\right.
\end{equation}
When the mass of the He layer reaches a certain value, He is assumed
to be ignited. If He-shell flashes occur, a part of the envelope
mass is assumed to be blown off. The mass-growth rate of WDs in this
case is linearly interpolated from a grid computed by Kate \&
Hachisu (2004), where a wide range of WD mass and mass-accretion
rate were calculated in the He-shell flashes. We define the
mass-growth rate of the CO WD, $\dot{M}_{\rm CO}$, as
 \begin{equation}
 \dot{M}_{\rm CO}=\eta_{\rm He}\dot{M}_{\rm He}=\eta_{\rm He}\eta_{\rm
 H}|\dot{M}_{\rm acc}|,
  \end{equation}
where $\eta_{\rm He}$ is the mass-accumulation efficiency for
He-shell flashes.

The evolution of these WD binaries is driven by the nuclear
evolution of the donor stars, and the change of the orbital angular
momentum of the binaries is mainly caused by wind mass-loss from the
WD. We assume that the mass lost from these binaries carries away
the same specific orbital angular momentum of the WD (the mass-loss
in the donor's wind is supposed to be negligible, but its effect on
the change of the orbital angular momentum, i.e. magnetic braking,
is included). We incorporate the prescriptions above into Eggleton's
stellar evolution code and follow the evolution of the WD + RG
systems. We have calculated a large, dense model grid, in which the
lobe-filling star is a RG star.

%\begin{figure*}
%\centerline{\epsfig{file=fig1a.ps,angle=270,width=8cm}\ \
%\epsfig{file=fig1b.ps,angle=270,width=8cm}} \caption{An example of
%binary evolution calculations. In panel (a), the solid and
%dash-dotted curves show $\dot M_2$ and $M_{\rm WD}$ varying with
%time, respectively. The open circles represent the evolution of
%$\dot M_{\rm CO}$ during outbursts. In panel (b), the evolutionary
%track of the donor star is shown as a solid curve and the evolution
%of orbital period is shown as a dash-dotted curve. Dotted vertical
%lines in both panels and asterisks in panel (b) indicate the
%position where the WD is expected to explode as a SN Ia. The initial
%binary parameters and the parameters at the moment of the SN Ia
%explosion are also given in these two panels.}
%\end{figure*}

\section{BINARY EVOLUTION RESULTS} \label{3. BINARY EVOLUTION RESULTS}
\subsection{An example of binary evolution calculations}

\begin{figure}
\includegraphics[width=50mm,angle=270]{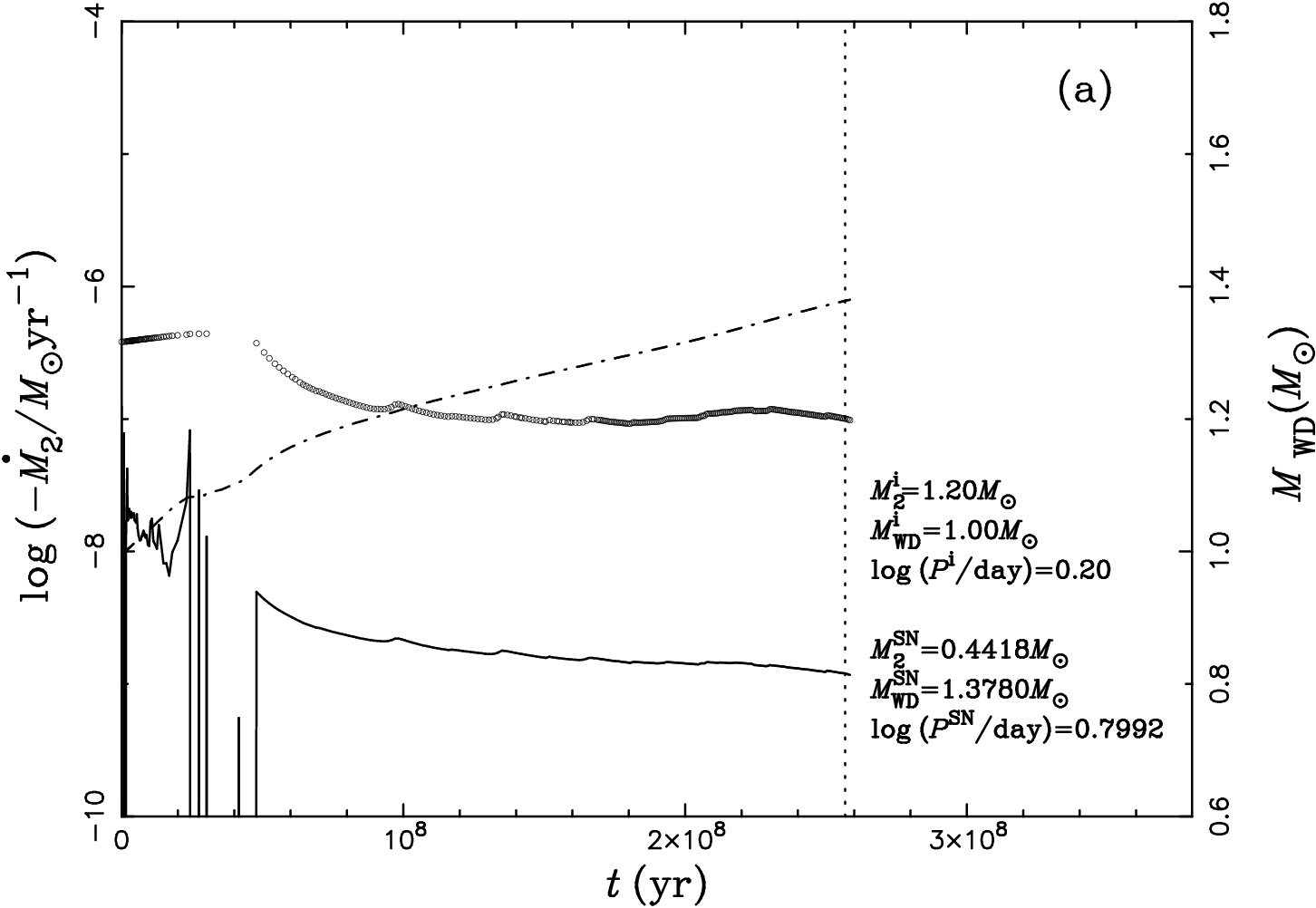}
\includegraphics[width=50mm,angle=270]{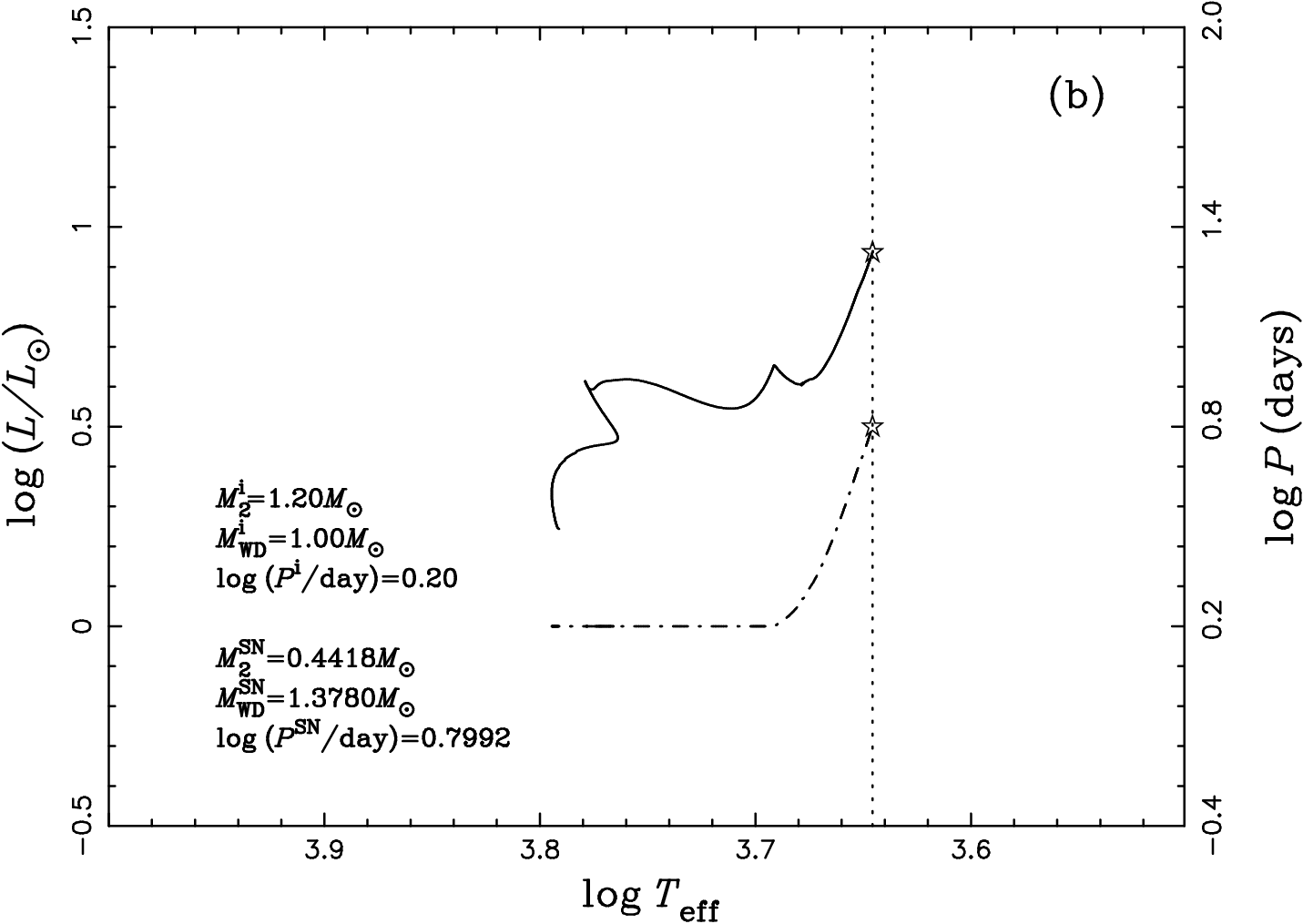}
\caption{An example of binary evolution calculations. In panel (a),
the solid and dash-dotted curves show $\dot M_2$ and $M_{\rm WD}$
varying with time, respectively. The open circles represent the
evolution of $\dot M_{\rm CO}$ during outbursts. In panel (b), the
evolutionary track of the donor star is shown as a solid curve and
the evolution of orbital period is shown as a dash-dotted curve.
Dotted vertical lines in both panels and asterisks in panel (b)
indicate the position where the WD is expected to explode as a SN
Ia. The initial binary parameters and the parameters at the moment
of the SN Ia explosion are also given in these two panels.}
\end{figure}

In Fig. 1, we present an example of binary evolution calculations.
Panel (a) shows the $\dot M_2$, $\dot M_{\rm CO}$ and $M_{\rm WD}$
varying with time, while panel (b) is the evolutionary track of the
donor star in the Hertzsprung-Russell diagram, where the evolution
of the orbital period is also shown. The binary shown in this case
is ($M_2^{\rm i}$, $M_{\rm WD}^{\rm i}$, $\log (P^{\rm i}/{\rm
day})$) $=$ (1.2, 1.0, 0.2), where $M_2^{\rm i}$, $M_{\rm WD}^{\rm
i}$ and $P^{\rm i}$ are the initial mass of the donor star and the
CO WD in solar masses, and the initial orbital period in days,
respectively. The donor star fills its Roche lobe on the RG stage
which results in case B mass-transfer. During the whole evolution,
the mass-transfer rate is lower than the critical value given by
Equation (1). Thus, the accretion disk experiences instability. The
mass-accretion rate $\dot M_{\rm acc}$ of the WD exceeds
$\frac{1}{8}\dot{M}_{\rm cr,WD}$ after the onset of RLOF, leading to
the mass-growth of the WD during outbursts. After about
$2.6\times10^{8}$\,yr, the WD grows to 1.378$\,M_{\odot}$, which
explodes as an SN Ia. At the SN explosion moment, the mass of the
donor star is $M^{\rm SN}_2=0.4418\,M_{\odot}$ and the orbital
period $\log (P^{\rm SN}/{\rm day})=0.7992$.

\subsection{Initial and final parameters for SNe Ia}
\begin{figure}
   \begin{center}
\includegraphics[width=6.cm,angle=270]{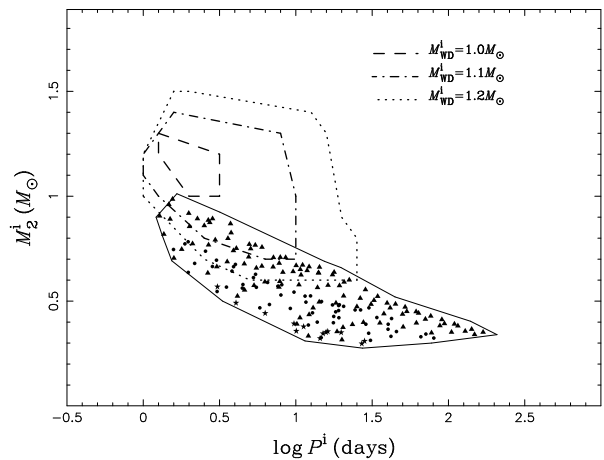}
 \caption{Parameter regions producing SNe
Ia in the orbital period--secondary mass plane ($\log P^{\rm i}$,
$M^{\rm i}_2$) for the WD + RG systems with the duty cycle $d=0.01$.
The initial WD masses are 1.0, 1.1 and 1.2$\,M_{\odot}$. The final
state of the WD + RG systems in the plane is encircled by the solid
line, where filled stars, circles and triangles denote that the WD
explodes as an SN Ia with initial WD masses of 1.0, 1.1 and
1.2$\,M_{\odot}$, respectively.}
   \end{center}
\end{figure}

In Fig. 2, we show the initial contours for SNe Ia and the final
state of binary evolution in the ($\log P^{\rm i}$, $M^{\rm i}_2$)
plane at the moment of SN Ia explosion with the duty cycle $d=0.01$.
The initial WD masses are 1.0, 1.1 and 1.2$\,M_{\odot}$. The final
contour (solid line in Fig. 2) is much lower than that of the
initial contours, which results from the mass-transfer from the
secondary to the WD and the mass-loss from the system. Note that,
the enclosed region almost vanishes for $M_{\rm WD}^{\rm
i}=1.0\,M_{\odot}$, which is then assumed to be the minimum WD mass
for producing SNe Ia from this channel. If the initial parameters of
a WD + RG system are located in the initial contours, an SN Ia is
then assumed to be produced. These initial contours for the WD + RG
channel can be expediently used in the binary population synthesis
studies. The data points and the interpolation FORTRAN code for
these contours can be supplied on request by contacting BW.

\section{DISCUSSION} \label{6. DISCUSSION}

\begin{figure}
   \begin{center}
\includegraphics[width=6.cm,angle=270]{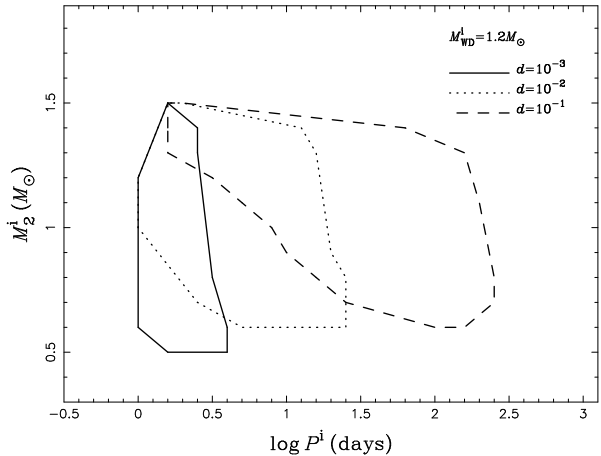}
 \caption{Regions in the initial orbital period--secondary mass plane ($\log P^{\rm i}$,
$M^{\rm i}_2$) for the WD + RG channel that produce SNe Ia with
initial WD mass of 1.2$\,M_{\odot}$, but for different duty cycle
values.}
   \end{center}
\end{figure}

Compared with most of previous investigations on the WD + RG channel
of SNe Ia, the main difference is that we considered the effect of
the accretion disk instability on the evolution of WD binaries. In
our results, there is no WD + RG system with period as long as
$\sim$$10^{\rm 3}$ days as indicated in Li \& van den Heuvel (1997)
and Hachisu et al. (1999a). This is because if the initial period of
a WD binary is too long, the mass-transfer rate between the RG donor
star and the WD will be too high and the optically thick wind will
occur and take much H-rich material away from the binary system.
Finally, the RG donor star has no enough material to be accumulated
onto the WD.

Similar to previous studies (e.g. King et al. 2003; Xu \& Li 2009),
our study also indicates that the long-period dwarf novae offer a
possible ways for producing SNe Ia. In particular, there is an
advantage for this work, i.e. the SN Ia explosion is always occur in
a WD binary of small secondary/primary mass ratio ($<$1), and with
very little of the H envelope of the secondary remaining in our
investigations. This feature will greatly reduce the possibility of
H contamination of the SN Ia ejecta (see also King et al. 2003),
which is consistent with the defining characteristic of most SNe Ia
having no detectable H (Branch et al. 1995). This also has broad
implications towards the work that is currently placing very tight
limits on the amount of H entrained in the SN Ia ejecta, determined
through the analysis of late-time spectra (e.g. Mattila et al. 2005;
Leonard 2007).

However, the results in this paper depend on many uncertain input
parameters, in particular for the duty cycle which is poorly known.
The main uncertainties lie in the facts that it varies from one
binary system to another and may evolve with the orbital periods and
mass-transfer rates (e.g. Lasota 2001; Xu \& Li 2009). This is the
reason why we choose an intermediate value (0.01) rather than other
extreme values (e.g. 0.1 or $10^{-3}$). Furthermore, we also did
some tests for a higher or lower value of the duty cycle. In Fig. 3,
we show the influence of the variation of the duty cycle value on
the regions for producing SNe Ia with initial WD mass of
1.2$\,M_{\odot}$. We see that, for a high value (0.1), the right
boundaries of the regions will be shifted to higher period, while,
for a low value ($10^{-3}$), the right boundaries of the regions
will be shifted to lower period. For the low value of the duty
cycle, it will have a high mass-accretion rate of WDs during
outbursts, so that the accreting WDs will lose too much mass via the
optically thick wind, preventing them increasing masses to the Ch
mass. Thus, a low value of the duty cycle will reduce the regions
for producing SNe Ia.

In this paper, we set the metallicity $Z=0.02$. For the WD + RG
channel, varying the metallicity would have strong influence on the
regions for producing SNe Ia (Fig. 2), e.g. high metallicity leads
to larger radii of zero-age MS (ZAMS) stars, then the left
boundaries of the regions will be shifted to longer period.
Meanwhile, stars with high metallicity evolve in a way similar to
those with low metallicity but less mass (Umeda et al. 1999; Chen \&
Tout 2007; Meng et al. 2009). Thus, for the WD binary systems with
particular orbital periods, the companion mass increases with
metallicity.

At present, the existence of a population of single low-mass
($<$0.45$\,M_{\odot}$) WDs (LMWDs) is supported by some observations
(e.g. Maxted et al. 2000; Kilic et al. 2007). The formation of
single LMWDs is still unclear. It is suggested that single LMWDs
could result from single old metal-rich stars which experiences
severe mass loss prior to the He flash (Kalirai et al. 2007; Kilic
et al. 2007). However, the study of initial-final mass relation for
stars by Han et al. (1994) implies that only LMWDs with masses
$\ga$0.4$\,M_{\odot}$ might be produced through such a single-star
channel, even at high metallicity (Meng, Chen \& Han 2008). Thus, it
would be difficult to conclude that single stars can produce the
LMWDs.

The companion in the WD + RG channel would survive and evolve to a
WD finally. In Fig. 2, we see that the companion stars have a low
mass at the moment of SN explosion. The companion stars will be
stripped of some mass due to the impact of SN ejecta. Marietta et
al. (2000) presented several high-resolution two-dimensional
numerical simulations of the impact of SN Ia explosion with
companions. They find that a RG donor star will lose almost its
entire envelope (96\%$-$98\%) owing to the impact of the SN Ia
explosion and leave only the core of the star. Thus, the surviving
companion stars from this channel will have a relatively low mass
after SN explosion and evolve to a WD finally, which provides a
possible pathway for the formation of the population of single LMWDs
($<$0.45$\,M_{\odot}$). Meanwhile, we also suggest that the
observed, apparently single LMWDs may provide evidence that at least
some SN Ia explosions have occurred with non-degenerate donor stars
(such as RG donor stars).

The CO WD + RG systems can be formed by binary evolution. Wang, Li
\& Han (2010) find that there is one channel which can form CO WD +
RG systems and then produce SNe Ia. In the detailed binary evolution
procedure, the primordial primary first fills its Roche lobe at the
thermal pulsing asymptotic giant branch stage. A common envelope is
then easily formed owing to dynamically unstable mass-transfer
during the RLOF stage. After the common envelope ejection, the
primordial primary becomes a CO WD, then a CO WD + MS system is
produced. The MS companion star continues to evolve until the RG
stage, i.e. a CO WD + RG system is formed. For the CO WD + RG
systems, SN Ia explosions occur for the ranges $M_{\rm
1,i}\sim5.0$$-$$6.5\,M_\odot$, $M_{\rm
2,i}\sim1.0$$-$$1.5\,M_\odot$, and $P^{\rm i}\ga 1500$\,days, where
$M_{\rm 1,i}$, $M_{\rm 2,i}$ and $P^{\rm i}$ are the initial mass of
the primary and the secondary at ZAMS, and the initial orbital
period of a binary system.

The WD + RG channel has a long delay time from the star formation to
SN explosion due to the RG donor star with low initial masses
($\la$1.5$\,M_{\odot}$). Thus, this channel can contribute to the
old population of SNe Ia implied by recent observations (Mannucci et
al. 2006; Totani et al. 2008; Schawinski 2009). The old population
of SNe Ia may have an effect on models of galactic chemical
evolution, since they would return large amounts of iron to the
interstellar medium much later than previously thought. It may also
have an impact on cosmology, as they are used as cosmological
distance indicators.

\section*{Acknowledgments}
We thank the anonymous referee for valuable comments that helped us
to improve the paper. This work is supported by the National Natural
Science Foundation of China (Grant No. 10821061), the National Basic
Research Program of China (Grant No. 2007CB815406), and the Yunnan
Natural Science Foundation (Grant No. 08YJ041001).

\label{lastpage}
\end{document}